\documentstyle[12pt]{article}    
\begin{document}
\begin{center}
   
      {\bf Semiclassical gravitation and quantization for the Bianchi type I universe with large anisotropy}

\vspace{1cm}

                       Wung-Hong Huang\\
                        Department of Physics\\
                         National Cheng Kung University\\
                         Tainan, 70101, Taiwan
\end{center}

 	We use a perturbative method to evaluate the effective action of a free scalar field propagating in the Bianchi type I spacetime with large space anisotropy.   The zeta- function regularization method is used to evaluate the action to the second order in the Schwinger perturbative formula.   As the quantum corrections contain fourth derivative in the metric we apply the method of iterative reduction to reduce it to the second-order form to obtain the self-consistent solution of the semiclassical gravity theory,   The reduced Einstein equation shows that the space anisotropy, which will be smoothed out during the evolution of universe, may play an important role in the dynamics of early universe.   We quantize the corresponding minisuperspace model to investigate the behavior of the space anisotropy in the initial epoch.   From the wavefunction of the Wheeler-DeWitt equation we see that the probability for the Bianchi type I spacetime with large anisotropy is less then that with a small anisotropy.

\vspace{3cm}
\begin{flushleft}

Classification Number: 04.60+v;11.10.gh\\
 E-mail: whhwung@mail.ncku.edu.tw\\
Physical Review D58 (1998) 084007
\end{flushleft}

{\newpage} 

\section  {INTRODUCTION}

 	The effective action plays an important role in investigating the theory of quantum fields in curved spacetime.   In the conformally flat spacetime the effective action can be completely determined by the local gometry [1].   If the spacetime is not conformally flat then the effective action could not be evaluated exactly and one can only evaluate it in perturbation.   Many years ago, Hartle and Hu [2] had used the dimensional regularization method to expand the effective action of a scalar field propagating in Bianchi type-I spacetime to the second order of space anisotropy.  From the effective action they investigated the problem of dissipating the space anisotropy by the quantum field effects [3].   This method has also been used to investigate the quantum field in an inhomogeneous spacetime [4] and a cosmic string [5].   However, as these results are restricted in the spacetime with small anisotropy it is of valuable to investigate the case with large anisotropy. 

 	In this paper a simple prescription for the expansion of effective action in the Bianchi type-I universe with large anisotropy is described.   We uses the zeta-function regularization method to evaluate the renormalized effective action to the second order in the Schwinger perturbative formula. [6].   As the quantum corrections contain up to fourth derivative in the metric the associated Einstein equation will suffer the problems of changing the Hamiltonian structure, lacking the stability, and appearing the unphysical solutions, etc. [7-9].   Therefore, we apply the method of iterative reduction to reduce it to a second-order equation [9].   We then investigate the reduced equation and see that the space anisotropy, which will be smoothed out during the evolution of universe, may play an important role in the early Universe.   To see the behavior of the space anisotropy in the initial epoch we quantize the corresponding minisuperspace model.   We analyze the Wheeler-DeWitt equation and see that probability for the Bianchi type-I spacetime with large anisotropy is less then that with a small anisotropy. 

 	This paper is organized as follows.   In section II a simple prescription to expand effective action in the Bianchi type-I universe with large anisotropy is described.   It then uses the zeta-function regularization method to evaluate the effective action to the second order in the Schwinger perturbative formula.   In section III the method of iterative reduction is used to reduce the action to the second-order form.   The solution of the reduced Einstein equation is then analyzed.   In section IV we quantize the reduced action and analyze the associated Wheeler-DeWitt equation.   The last section is devoted to a short summary.

\section  {EXPANSION  OF  EFFECTIVE  ACTION  WITH  LARGE ANISOTROPY}

\subsection  {Method}
 	We consider the Lagrangian describing a massless scalar field conformally coupling to the gravitational background

$$L = - {1\over2}g^{\mu\nu}\partial_\mu\Phi\partial_\nu\Phi -{1\over12}R\Phi^2, \eqno{(2.1)}$$
where R is the curvature scalar.   We will calculate the renormalized effective action in the Bianchi type-I spacetime with the line element [10]

     $$d^2s = g_{\mu\nu}dx^{\mu}dx^{\nu} =- dt^2 + \sum_{i}e^{2\beta_i}dx_i^2,\eqno{(2.2)}$$
where

 $$ \beta_i=\beta_{+} + \sqrt{3}\beta_{-},   ~~~~~  \beta_2=\beta_{+} - \sqrt{3}\beta_{-}, ~~~~~ \beta_3=-2\beta_{+}.
\eqno{(2.3)}$$
The effective action in the metric  ${\tilde g}_{\mu\nu}=a(t)^{2} g_{\mu\nu}$ can be found through a conformal transformation formula, which will be described in II.3.  

   The Hamiltonian associated with the Lagrangian described in (2.1) can be written as 

		$$H =H_0 + D + (Q - Q_0) .                                    \eqno{(2.4)}$$
where
	$$H_0= \partial_t^2 - \sum_{i}\partial_{x_i}^2 + Q_0 ,                                    \eqno(2.5)$$
       $$ D= \sum_{i}D_i \partial_{x_i}^2  =  \sum_{i} (1 - e^{-2\beta_i}) \partial_{x_i}^2,\eqno(2.6) $$
and
 		$$Q=\dot\beta^2_{+} + \dot\beta^2_{-}  \eqno(2.7)$$
denotes the quantity of the space anisotropy.  

   In this paper we will investigate the back-reaction effect only for the universe near the initial epoch.   If we denote a large space anisotropy at initial time t = 0 by $ Q_0$  then the quantity $(Q - Q_0)$ in (2.4) will be very small if the universe is sufficiently near the initial epoch.   From (3.11) and (3.13) we see that for the universe near the initial epoch the anisotropy $Q_0$ can be very large but the function $\beta_j$ and the quantity  $D_j$  are very small.   Thus the effective action can be evaluated in perturbation by regarding $D_j$ and$ (Q - Q_0)$ as the small quantities   The fact that $D_j$ be small is very crucial for the perturbation expansion used in this paper.   In short,  the limitation of the computation in this paper is that it is valid only for a very short time; indeed, for a parametrically short time.   It is in this limitation that the $\beta_j$ and the quantity  $D_j$ are small and the derivative of $\beta_j$  (and thus the anisotropy $Q_0$) are large at the initial time.    Thus we can regard $D_j$ and $(Q - Q_0)$ as the small quantities and the effective action can be evaluated in perturbation.

\subsection  {Calculations}
   We will evaluate the renormalized effective action by the $\zeta$-function regularization [1,11] method 

		$$W = -{i\over2}  ln[Det(H)] = -{i\over2} [\zeta'(0) + \zeta(0) ln(\mu^2)].             \eqno(2.8) $$
In the proper-time formalism, the $\zeta$ function can be found from the relation

$$\zeta(\nu) = [\Gamma(\nu)]^{-1} \int dx^4 [-g(x)]^{1/2} \int_0^{\infty}  ids (is)^{\nu -1}<x\mid e^{-isH}\mid x>,    \eqno(2.9)$$
where the operator H is defined in (2.4) and it is understood that $H \rightarrow H - i\epsilon$, with $\epsilon$ a small positive quantity.   As the quantities $(Q - Q_0)$ and $D$ are small the $\zeta$ function can be calculated by the Schwinger perturbative formula [6] 

 $$Tr e^{-isH} = Tr [ e^{-isH_0} - is e^{-isH_0} (Q - Q_0 )  - is e^{-isH_0} D +{s^2\over2}\int_0^{1} du e^{-is(1-u)H_0} (Q - Q_0) e^{-isuH_0} (Q - Q_0) $$
  $$+{s^2\over2}\int_0^{1} du e^{-is(1-u)H_0} (Q - Q_0) e^{-isuH_0} D +{s^2\over2}\int_0^{1} du e^{-is(1-u)H_0} D e^{-isuH_0} (Q - Q_0) $$ 
 $$+{s^2\over2}\int_0^{1} du e^{-is(1-u)H_0} D e^{-isuH_0} D + \cdot\cdot\cdot] .                 \eqno(2.10)$$
Then the $\zeta$ function can be expressed as

$$ \zeta(\nu)= \zeta(\nu)_0 +\zeta(\nu)_Q +\zeta(\nu)_D +\zeta(\nu)_{QQ} +\zeta(\nu)_{QD} +\zeta(\nu)_{DQ} +\zeta(\nu)_{DD} +\cdot\cdot\cdot, \eqno(2.11)$$
where $\zeta_ i(\nu)$and $\zeta_{ij}(\nu)$ are defined and calculated in below.

   We first calculate $\zeta_0(\nu)$  
	$$\zeta_0(\nu)= [\Gamma(\nu)]^{-1} \int dx^4 [-g(x)]^{1/2} \int_0^{\infty}  ids (is)^{\nu -1}<x\mid e^{-isH_0}\mid x>$$
	$$= [\Gamma(\nu)]^{-1} \int dx^4 \int{d^4p\over (2\pi)^4} \int_0^{\infty} ds (s)^{\nu -1} e^{-s(p^2+Q_0)} $$
	$$= i\int d4x \int{d^4p\over (2\pi)^4} (p^2 + Q_0)~~~~~~~~~~~~~~~~~~~~~~~~~~~$$
	$$= i (32\pi^2)^{-1}\int dx^4 ~Q_0^2 ~[ 1 + \nu \left({3\over2}  - lnQ_0 \right) + O(\nu^2)].   \eqno(2.12)$$
To obtain the above result we have carried the integration by letting $s \rightarrow i s$ and rotating $p_0$ through  ${\pi\over2}$  in the complex plane.  This procedure will also be used in the followings. 

   Next, we calculate $\zeta_Q(\nu)$

 	$$\zeta_Q(\nu)= [\Gamma(\nu)]^{-1} \int dx^4 [-g(x)]^{1/2} \int_0^{\infty}  ids (is)^{\nu -1}(-is) <x\mid e^{-isH_0}(Q-Q_0)\mid x>$$
$$= [\Gamma(\nu)]^{-1} \int dx^4 (Q-Q_0) \int{d^4p\over (2\pi)^4} \int_0^{\infty} ds s^{\nu} e^{-s(p^2+Q_)}~~~~~~~~~~~~~~~~~~~~$$
$$= i (16\pi^2)^{-1} \int dx^4 (Q - Q_0) Q_0 [1 + \nu (1 - lnQ_0 ) + O(\nu^2)].  {\hspace{2cm}}      \eqno(2.13)$$
In the same way we have 
 	$$\zeta_D(\nu)= [\Gamma(\nu)]^{-1} \int dx^4 [-g(x)]^{1/2} \int_0^{\infty}  ids (is)^{\nu -1}(-is) <x\mid e^{-isH_0}D\mid x>$$
$$= i (64\pi^2)^{-1} \int dx^4 (\sum_i D_i)  Q_0^2 [1 + \nu ({3\over 2} - lnQ_0 ) + O(\nu^2)].  {\hspace{2cm}}      \eqno(2.14)$$

 	We now tern to evaluate the functions $\zeta_{ij}(\nu)$ which involving more calculations.  From the definition we have 

$$\zeta_{QQ}(\nu)= \Gamma(\nu)^{-1} \int dx^4 [-g(x)]^{1/2} \int ids (is)^{\nu -1}(-{s^2\over 2})\int_0^1 du  <x\mid e^{-is(1-u)H_0}(Q-Q_0) e^{-isuH_0} (Q-Q_0)\mid x>$$
     $$=-(2\Gamma(\nu))^{-1}\int dx^4 [Q(x_0) - Q_0]\int dy^4 [Q(y_0) - Q_0]   \int{d^4p\over (2\pi)^4} \int{d^4k\over (2\pi)^4} \int ds (s)^{\nu +1}\times$$
$$\int du e^{- s(1-u)[k^2+Q_0(x_0)] +s u[p^2+Q_0(y_0)]} e^{i(p-k)\cdot(x-y)} ,  $$
in which we have inserted the complete set $\sum _k |k > <k|$ or $\sum_p |p > <p| $ before the operator $H_0$.   After integrating the variables $u$,$ s$, $y$, $p$, and $k$, then shifting $k_0 \rightarrow k_0  + p_0$ and performing the $p_0$, $k_0$ and $y_0$  integrations we have the final result

$$\zeta_{QQ}(\nu) = - i {\pi \over 4}\int dx^4 (Q-Q_0)^2 [1 - \nu lnQ_0 + O(Q_0^{-1}) + O(\nu^2)] .   \eqno(2.15)$$
Through the same manipulation we have

$$\zeta_{QD}(\nu)= \Gamma(\nu)^{-1} \int dx^4 [-g(x)]^{1/2} \int ids (is)^{\nu -1}(-{s^2\over 2})\int_0^1 du  <x\mid e^{-is(1-u)H_0}(Q-Q_0) e^{-isuH_0} D)\mid x>$$
$$= - i {3\pi \over 24}\int dx^4 (Q-Q_0)Q_0 \sum_j D_j[1 + \nu(1- lnQ_0) + O(Q_0^{-1}) + O(\nu^2)] .	    \eqno(2.16)$$

$$\zeta_{DQ}(\nu)= \Gamma(\nu)^{-1} \int dx^4 [-g(x)]^{1/2} \int ids (is)^{\nu -1}(-{s^2\over 2})\int_0^1 du  <x\mid e^{-is(1-u)H_0}D e^{-isuH_0}(Q-Q_0))\mid x>$$
$$= - i {3\pi \over 24}\int dx^4 (Q-Q_0)Q_0 \sum_j D_j[1 + \nu(1- lnQ_0) + O(Q_0^{-1}) + O(\nu^2)] .	    \eqno(2.17)$$

$$\zeta_{DD}(\nu)= \Gamma(\nu)^{-1} \int dx^4 [-g(x)]^{1/2} \int ids (is)^{\nu -1}(-{s^2\over 2})\int_0^1 du  <x\mid e^{-is(1-u)H_0}D e^{-isuH_0} D)\mid x>$$
	$$= - i {\pi \over 16}\int dx^4 [2(\sum_j D_j)^2+\sum_j D_j^2][1 + \nu ({3\over 2} -lnQ_0)] + O(Q_0^{-1}) + O(\nu^2)]. \eqno(2.18)$$

  	Now, substituting the calculated $\zeta$ functions into (2.8) we finally obtain the effective action $W[g_{\mu\nu}]$ for the free scalar field propagating in the metric (2.2).   

$$	W[g_{\mu\nu}] = \int dx^4 [- {\pi^2\over 64}  Q_0^2 lnQ_0 - {\pi^2\over 32}  (Q - Q_0) Q_0 lnQ_0  + {\pi^2\over 128}(\sum_j D_j ) Q_0^2 lnQ_0 ], 	    \eqno(2.19)$$
in which, for convenience, we have let the parameter $\mu = 1$.   The effective action in (2.19) contains only these from $\zeta_0$ , $\zeta_Q$and $\zeta_Q$.   Those from $\zeta_{ij}$ is smaller then these from $\zeta_i$  and are neglected therefore.   The Lagrangian density coming from $\zeta_0$ is a constant and can be absorbed by the renormalization of cosmological constant. 

 	Note that the effective action in (2.19) dose not have imaginary part and thus there has no particle been produced [1].   This is because that in our approximation we assume that both of $(Q-Q_0)$ and $D$ are small.   This means that we constrain the system near the initial epoch, thus the universe has yet not evolved too much and the particle has yet not been produced.   This behavior had also been found in our previous investigation about a rotational spacetime [12].   Following the above calculations we can see that if we let the universe evolve , thus $Q$ approaches zero, then the effective action will have imaginary part and there dose have particle been produced.   However, in this case the universe will far from the initial stage.   However, as will be seen in the next section , what we are concerning with is the state near the initial epoch, not that far from the initial epoch.   Thus, the effective action in (2.19) is a suitable one which will be used to analyze in the next section.   Note also that the imaginary part, if it appears, may possibly be shown in $\zeta_{ij}$ , as those in [12,13].   Thus, in this section we calculate the zeta function to the second order in the Schwinger perturbative to see whether the imaginary part will be appear.   However, our calculation shows that it dose not appear, at least to the order we have evaluated.

\subsection   {Conformal transformation}
	The more cosmologically interesting spacetime is that described the metric
$$		ds^2 = \tilde g_{\mu\nu}dx^\mu dx^\nu = a(t)^2  g_{\mu\nu}dx^\mu dx^\nu = a(t)^2 [- dt^2 + \sum_i e^{2\beta_i} dx_i^2 ] . \eqno(2.20)$$
Now the effective action in the metric $\tilde g_{\mu\nu} = a(t)^2 g_{\mu\nu}$ , denoted as $W [\tilde g_{\mu\nu}]$ , can be easily found by the conformal transformation formula [13]

$$W [\tilde g_{\mu\nu}]= W[g_{\mu\nu}] +{1\over 16 \pi^2} \hbar  \int dx^4 [-g(x)]^{1/2}[ ~A[U(R_{\mu\nu\lambda\delta} R^{\mu\nu\lambda\delta}- 4 R_{\mu\nu} R^{\mu\nu}+ R^2) + 2 RU,_\mu U,^\mu  - 4R_{\mu\nu} U,^\mu U,^\nu$$

$$  - 4 U,_\mu^{;\mu} U,_\lambda U,^\lambda  - 2 (U,_\lambda U'^\lambda )^2 ] + B [U(R_{\mu\nu\lambda\delta} R^{\mu\nu\lambda\delta}- 2 R_{\mu\nu} R^{\mu\nu}+{1\over 3} R^2) +$$
$$ {2\over 3} R(U,_\mu^\mu + U,_\mu U,^\mu ) -  2U,_\mu U,^\mu - 4 U,_\mu^{;\mu} U,_\lambda U,^\lambda  - 2(U,_\lambda U,^\lambda)^2 ]~ ],          \eqno(2.21)$$
where $U= ln(a)$ , $A = - 1/360 $ and $B = 1/120$.    This formula also appears in [14] in which the other applications were discussed.   The explicit expression of (2.21) will be shown in next section.

\section  {REDUCED  ACTION  AND  SEMICLASSICAL  GRAVITATION }

 	From the effective actions shown in (2.19) and (2.21) we see that they contain fourth derivative in the metric.   The associated system will thus suffer the problems of changing the Hamiltonian structure, lacking the stability, and appearing the unphysical solutions, etc. [7-9].   Therefore, we will apply the method of iterative reduction to reduce it to a second-order equation.   We will follow the method which used by Parker and Simon [9] to find the reduced system for the spacetime without space anisotropy.
	The method of iterative reduction is as follows.  The semiclassical gravitation theory is that described by the generalized Einstein equation with quantum corrections
	$$ R_{\mu\nu} -{1\over 2 } g_{\mu\nu}  R = 8\pi G [ T_{\mu\nu} ^c + T_{\mu\nu}^q ] ,  \eqno(3.1)$$
where $T_{\mu\nu}^c$  is the stress tensor of the classical radiation which is included to support the expansion of the Universe at the late epoch compared with the Planck time.   Other matter is neglected because it makes negligible [9].  The quantity $T_{\mu\nu}^q$  is the renormalized stress tensor of the quantum field, which can be found from the renormalized effective action $W [\tilde g {\mu\nu}]$ .   Because $T_{\mu\nu}^q$  is the quantum correction term, it will be first order in $\hbar$.   Thus, to the zero order of $\hbar$, we can neglect  $T_{\mu\nu}^q$ .

   Therefore
 	(I) From the Bianchi identity ${T^c_{\mu\nu}} ^{;\nu} = 0$ we have a relation 
             $$   \epsilon \approx \epsilon_0 a^{-4} ,	 \eqno(3.2)$$
 where $ \epsilon $ is the energy density of classical radiation and $ \epsilon_0 $ an integration constant. 

 	(II) The spatial-spatial component Einstein equation leads to

 $$ a^{-2}(\dot a ^2 a^{-2}+ \ddot  a  a^{-1} + \ddot\beta_ i+ 2 \dot \beta_ i \dot a  a^{-1} ) = {8\pi G\over 3}\epsilon .    \eqno(3.3)$$
It then implies two useful relations  

$$	a^{-2}(\dot a ^2 a^{-2}+ \ddot a   a^{-1}) = {8\pi G\over 3}\epsilon,  \eqno(3.4)$$
$$	\ddot \beta_{\pm} + 2 	\dot \beta_{\pm} a^{-1}  = 0 , ~~~~ \Rightarrow \dot \beta_{\pm}\approx c_{\pm}  a^{-2}  ,		 \eqno(3.5)$$
where $c _{\pm}$ are the integration constants.

   (III) The time-time component Einstein equation leads to

  $$   \dot a ^ 2  a^{-2} - Q = {8\pi G\over 3} a^2 \epsilon.  \eqno(3.6)$$
Then from Eqs.(3.2),(3.5) and (3.6) we have the relation

 $$		\dot a ^ 2 \approx {8\pi G\over 3}\epsilon _0 + A a^{-2},  \eqno(3.7)$$
where 
 $$ 		A = (c_+^2 + c_-^2 ) .	 \eqno(3.8)$$
Differentiating Eq.(3.7) leads to
 $$ 		 \ddot   a  = - A a^{-3} .			 \eqno(3.9)$$
And differentiating Eq.(3.9) leads to
 $$ 		\dot{} \ddot   a  = 3 A a^{-4} \dot a	.		\eqno(3.10)$$
These relations will be substituted into the renormalized effective actions (2.19) and (2.21) to reduce them to second-order equations.  Before doing this we will see two useful results.

 	 First, from (3.5) we see that that the space anisotropy defined in (2.7) becomes

 $$ 		 Q = A a^{-4}. 			\eqno(3.11)$$
Thus the anisotropy will be smoothed out during the evolution of expanding Universe (i.e., if "a" becomes large.) .   However, if the value of  "A", which is proportional to the space anisotropy Q, was not too small then the anisotropy may be very large when "a" is small (i.e., near the initial epoch).   This means that the anisotropy may affect the evolution of the Universe in the early epoch.   However, it is known that the determination of the constant "A" in a theory is the question of the initial problem and it can only be solved by the theory of the quantum cosmology [15.16].   Thus, we will quantize the reduced semiclassical gravitation theory to investigate this problem in the next section.

   Next, from (3.7) it is seen that when  "a"  is small then we can neglect the $\epsilon_0$ term and it has a solution 

         $$  a = (2At + a_0^2)^{1/2} . 	              \eqno(3.12)$$
Substituting this relation into (3.5) we find that  
 $$ \beta_j = c_j (2A)^{-1} ln[(2At + a_0^2) a_0^{-2}]. \eqno(3.13a) $$
Thus, $\beta_j$  become mall as $t \rightarrow  0$.   In this case we see that the functions $D_j$ defined in (2.6) becomes 
 $$			D_j = 2c_j (2A)^{-1}~ln[(2At + a_0^2) a_0^{-2}] ,	   \eqno(3.13b) $$
which can be regarded as a small function as $t \rightarrow  0$.   This proves the crucial assumption adopted in the section 2-1.   Let us emphasize that the relations of (3.12) and (3.13) can only be used for the universe near the initial epoch and  our calculation in the section 2-2 is a good approximation only for the universe near the initial time. 

   Before entering into the next section we will mention that the constant $a_0$ in (3.12) will not be zero.   This is because that as $a \rightarrow 0$  the time-time component of the reduced Einstein equation becomes 

  $$  a^{-2} \dot a^ 2 + {4\over 9} A^2 a^{-6} - A a^{-4} - {8\pi G\over 3} \epsilon_0 a^{-2} = 0 ,\eqno(3.14)$$
in which the first term coming from the classically Einstein action, the second and third terms are coming from the leading part of the quantum corrections in (2.21) and (2.19) respectively, while the last term is that from the classical radiation.   The Eq.(3.14) shows an interesting fact that the classical radiation dose not affect much the initial state of the universe and it is the conformal part that will dominate the contribution.   Now Eq.(3.14) could be regarded as a equation describing a particle with a unit mass in a potential U(a) which become positively infinite as  $a \rightarrow 0$ and thus the universe will be bounced at a positive radial $a_0$.   Note that the radial $a_0$ must be a small value because the conformal part is a quantum correction which will contain $\hbar$.   Such a bounce solution is also shown in the semiclassical gravitation model in the spacetime without anisotropy, as analyzed by Parker and Simon [9].

\section  {QUANTIZATION  OF  THE  REDUCED  ACTION }

Before performing the quantization we need some manipulations.   First, for the later convenience we will change the metric to be
$$ 		ds^2 = - dt^2 + e^{2\alpha}\sum_i  e^{2\beta_i }dx_i^2 ,	 		 \eqno(4.1)$$
which is a conventional form used in the Bianchi type-I quantum cosmology [17,18].   In this metric form, after the calculation, we can from (2.19) and (2.21) find that 

$$ 	W[g_{\mu\nu}] = \int dx^4  e^{-\alpha} [- {\pi^2\over 64}  Q_0^2 ~lnQ_0 - {\pi^2\over 32} (Q - Q_0) Q_0 ~lnQ_0  +{\pi^2\over 128} (\sum_j D_j) Q_0^2 ~lnQ_0] .     \eqno(4.2)$$

$$  W[\tilde g_{\mu\nu}] = W[g_{\mu\nu}] + {1\over 1920\pi^2}\hbar \int dx^4  e^{3\alpha}[{2\over 3} \dot \alpha^4  - 2(\ddot \alpha  + 2 \dot \alpha^2)^2 + 48 \alpha Q^2 - 4 (\ddot \alpha + 3 \dot \alpha ^2)Q$$
$$ + 12 \alpha [\ddot \beta_+^2 + \ddot \beta_-^2 +2 \dot \alpha (\ddot \beta_+  \dot \beta_-+ \dot \beta_+  \ddot \beta_-) + \dot\alpha^2 Q] +  {32\over 3}\dot \alpha \dot\beta^3 -32 \dot\alpha \dot\beta_+\dot\beta_-^3] .              \eqno(4.3)$$
Next, we express the reduction relations found in (3.2), (3.5), (3.7), (3.8), (3.9) and (3.10) in the metric form (4.1), and then substitute these new relations into (4.2) and (4.3).   The results are 

$$W[g_{\mu\nu}]= \int dx^4  e^{-\alpha} [- {\pi^2\over 64} A^3 a_0^{-8}~ ln(A a_0^{-4}) - {\pi^2\over 32} A^2(a^{-4} - a_0^{-4})a_0^{-4}~ln(A a_0^{-4})$$
$$+{\pi^2\over 128} (\sum_j D_j) A^2 a_0^{-8}~ ln(A a_0^{-4}) ] . \eqno(4.4)$$

$$  W[\tilde g_{\mu\nu}] = W[g_{\mu\nu}] + {1\over 1920\pi^2}\hbar \int dx^4  e^{3\alpha}[{2\over 3}({8\pi G\over 3}  e^{-4\alpha} + A e^{-6\alpha})^2 - 2({32\pi G\over 3}  e^{-4\alpha} - A e^{-6\alpha})^2$$
$$  + 96 \alpha  A^2 e^{-12\alpha } - {160\pi G\over 3}A e^{10 \alpha} + 32 \alpha  e^{-6 \alpha}( 2 c_+^2  A + 4 c_+ c_-^3)({8\pi G\over 3}  e^{-4\alpha} + A e^{-6\alpha})^{1/2}].    \eqno(4.5)$$
Finally, the classical action $W_c [g_{\mu\nu}]$ which shall be added before analyzing the quantization of reduced system is 
$$ W_c [\tilde g_{\mu\nu}] = {1\over 16\pi G}\int dx^4 e^{3\alpha} [ - 6\dot \alpha^ 2 + 6(\dot \beta_+^2 +\dot \beta_-^2 )],   \eqno(4.6)$$
where G denotes the gravitational constant.
 	The total action to be quantized is the summation of  (4.4), (4.5) and (4.6).   Form (4.4) and (4.5) and we see that after using the method of iterative reduction the quantum correction dose not change the canonical momentums used in the classical action.   Thus the canonical momentums are 

 $$\Pi_\alpha = \delta W_c [\tilde g_{\mu\nu}] /\delta\alpha = - 6\dot \alpha  e^{3\alpha} ,  \eqno(4.7)$$
 $$\Pi_{\beta_\pm} = \delta W_c [\tilde g_{\mu\nu}] /\delta{\beta_\pm} =  6\dot {\beta_\pm}  e^{3\alpha} ,  \eqno(4.8)$$
which are those used in the quantum cosmology model without the quantum-field effects [17,18].   Then, the Hamiltonian is defined by 

$$ 	 H = \Pi_\alpha \dot\alpha  + \Pi_{\beta_+}\dot {\beta_+} + \Pi_{\beta_-}\dot {\beta_-} - L ,      \eqno(4.9)$$
in which the Lagrangian density $L$ can be found from the total action.   

    Now, through the canonically substituting
 $$	\Pi_\alpha \rightarrow- i \partial /\partial \alpha,			\eqno(4.10)$$
 $$     \Pi_{\beta_\pm}\rightarrow - i \partial /\partial \beta_\pm.		\eqno(4.11)$$
the Wheeler-DeWitt equation associated with the reduced semiclassical gravitational equation becomes
 
$$ {\partial^2 \Psi\over\partial\alpha^2}  + {\partial^2 \Psi\over\partial{\beta_+}^2}  + {\partial^2 \Psi\over\partial{\beta_-}^2} +  U(\alpha, A, \beta_+,\beta_- ) \Psi = 0, 		\eqno(4.12)$$
where the potential U is defined by
$$ U(\alpha, A, \beta_+,\beta_- ) = U_\alpha(\alpha) + U_\beta(\alpha, \beta_+, \beta_-) ,  \eqno(4.13)$$
in which
$$ 	 U_\alpha(\alpha) = {3\over 5\pi^2} A^2 e^{-6\alpha} |\alpha| , $$

$$	 U_\beta(\alpha, \beta_+, \beta_-)  = - {9\over8 \pi^2} e^{2\alpha}(\beta_+^2 + \beta_-^2) A^2 e^{-8\alpha_0} ~ ln(A e^{-4\alpha_0 }). 			\eqno(4.14)$$
To obtain the above equation we have let $8\pi G= 1$ and considered only the initial epoch, $\alpha \rightarrow \infty$, i.e., $ a \rightarrow  0$.    The Wheeler-DeWitt equation (4.12) is too complex to be solved exactly and some approximations shall be adopted.

 	 When the universe near initial epoch we can replace $\alpha$  in $U_\beta$  by $\alpha_0$ .   Then $U_\beta$ is only function of $\beta_+$ and $\beta_-$, and thus (4.12) could be separated into two equations

 $${\partial^2 \Psi_\alpha\over\partial\alpha^2} + ({3\over 5\pi^2} A^2 e^{-6\alpha} |\alpha|  - C)\Psi_\alpha = 0, \eqno(4.15)$$

$$ {\partial^2 \Psi_\beta\over\partial{\beta_+}^2}  + {\partial^2 \Psi_\beta\over\partial{\beta_-}^2}+[C- {9\over 8 \pi^2} e^{2\alpha}(\beta_+^2 + \beta_-^2) A^2 e^{-8\alpha_0} ~ ln(A e^{-4\alpha_0 })] \Psi_\beta= 0,\eqno(4.16)$$
if we take 

   $$\Psi = \Psi_\alpha\Psi_\beta ,	\eqno(4.17) $$
Note that the number C appearing in (4.15) and (4.16) can be neglected if $\alpha_0 \rightarrow - \infty$.

   Now, Eq.(4.15) can be solved in the WKB approximation and the solution is 

$$\Psi_\alpha  = |\alpha|^{-1/4} e^{-3|\alpha|/2} exp[- i{3A\over 5\pi} |G(\alpha)| ]   \eqno(4.18)$$
where
$$ 	G(\alpha) = \int  d\alpha |\alpha| e^{ -3\alpha} . 	\eqno(4.19)$$
Eq.(4.16) can be solved exactly [19] and we have 

 $$  \Psi_\beta = {K_s} [ i ~{3\over\pi} 8^{-1/2} Ae^{-3\alpha_ 0} [(\beta_+^2 +\beta_-^2) ~ ln(A e^{-4 \alpha_0 })]^{1/2}],  \eqno(4.20)$$
in which $s$ is an integration constant, which is irrelevant to our discussion in below,  and $K_s$ is a Bessel function.   From the asympotical behavior of the Bessel function [19] 

  $$ K_s(z) = ({\pi \over 2 z})^{1/2}~ e^{-z} [1 +O(z^{-1})] , 	\eqno(4.21)$$
we see that  $\Psi_\beta$  is a decreasing function of $A$ and $\Psi_\beta \rightarrow 0$  if $A  \rightarrow \infty$.   Because the probability density of universe is proportional to  $\mid\Psi_\alpha \Psi_\beta \mid^2$ ,   we thus conclude that the Bianchi type-I universe is not likely to be in a state with large space anisotropy.   These complete our investigations.

\section  {CONCLUSION}

We have presented a prescription to expand the effective action in the Bianchi type-I spacetime with large anisotropy.   We uses the zeta-function regularization method to evaluate the renormalized effective action of a quantum scalar field to the second order in the Schwinger perturbative formula.   As the quantum corrections contain up to fourth derivative in the metric, to obtain the self-consistent solutions of the semiclassical gravity theory, we apply the method of iterative reduction to reduce it to a second-order equation.   The reduced equation shows that the space anisotropy, which may play an important role in the early Universe, will be smoothed out during the evolution of Universe.   We thus quantize the corresponding minisuperspace model to investigate the behavior of space anisotropy in the initial epoch.   We solve the Wheeler-DeWitt equation in an approximation.   From the wavefunction of the Wheeler-DeWitt equation we see that the probability of the Bianchi type I spacetime with large anisotropy is less then that with a small anisotropy.   Thus, the Bianchi type-I universe is not likely to be in the state with large space anisotropy

  Finally, let us make two remarks.

     (1) In a previous paper [20] we had quantize the effective action with small anisotropy ( which was first evaluated by [2]) and see that universe is likely to be in the state with small anisotropy.   In this paper we quantize the effective action with large anisotropy (which is first evaluated in this paper) and see that universe is not likely to be in the state with large anisotropy.
 
    (2) It can be seen that the prescription used in this paper may also be applied to other cosmologically interesting spacetimes.   However it shall be mentioned that, although in general we can absorb some large quantities (in this paper it is $Q_0$) in initial epoch into $H_0$, but this dose not ensure that the remaining terms (in this paper they are $D_j$) are small.   Only if the chosen $H_0$ was sufficiently simple and remaining term was small could the perturbative method be useful.   The investigations about other spacetime, such as the Bianchi Type IX rotating universe, will be discussed elsewhere.

\newpage

\begin{enumerate}

\item  N. D. Birrell and P. C. W. Davies, Quantum Field in Curved Space (Cambridge University Press, Cambridge, England, 1982)
\item  J. B. Hartle and B. L. Hu, Phys. Rev. D 20, 1772 (1979).
\item   J. B. Hartle and B. L. Hu, Phys. Rev. D 21, 2756 (1980); J. B. Hartle, ibid, 22, 2091 (1980).
\item   W. H. Huang, Phys. Rev. D 48, 3914 (1993).
\item    A. Campos and E. Verdaguer, Phys. Rev. D 49, 1861 (1994). 
\item    J. Schwinger, Phys. Rev. 82, 664 (1951).
\item    L. Bel and H. S. Zia, D 32, 3128 (1985); J. Z. Simon, ibid. 41, 3720 (1990).
\item    F. D. Mazzitelli, Phys. Rev. D 45 (1992)
\item    L. Parker and J. Z. Simon, Phys. Rev. D 47, 1339 (1993).
\item    C. W. Misner, K. S. Thorne, and J. A. Wheeler, Gravitation (Freeman, San Francisco, 1973). 
\item    L. Parker, " Aspects of quantum filed theory in curved spacetime: effective action and energy-momentum tensor" in "Recent Developments in Gravitation", ed. S. Deser and M. Levey (New York: Plenum, 1977).
\item    W. H. Huang, One-loop Effective Action on Rotational Spacetimes: $\zeta$ - Function Regularization and Swinger Perturbative Expansion, Ann. Phys. 254, 69 (1997).
\item    W. H. Huang, Conformal Transformation of Rnormalized Effective in Curved Spacetimes,Phys. Rev. D 51, 579 (1995). 
\item    M. R. Brown and A. C. Ottewill, Phys. Rev. D 31, 3854 (1987); J. S. Dowker, ibid. 33, 3150 (1986); J. S. Dowker and J. P. Schofield, ibid. 38, 3327 (1988). 
\item    J. B. Hartle and S. Hawking, Phys. Rev. D 28, 2960 (1983)
\item    J. J. Halliwell, "Introductory lectures on quantum cosmology " in " Quantum cosmology and baby Universe", ed. S. Coleman, J. B. Hartle, T. Piran and S. Weinberg, (World Scientific Publishing Co. Pte. Lad. 1991), and references therein. 
\item    J. Louko, Ann. Phys. (N. Y.) 181, 318 (1988).
\item    S. W. Hawking and J. C. Luttrell, Phys. Lett B 143, 83 (1984); B. Berger and C. N. Vogeli, Phys. Rev. D 32, 2477 (1985); B. Berger , ibid. 32, 2485 (1985); K. Schleich, ibid. 39, 2192 (1989)
\item    I. S. Gradshteyn and I. M. Ryzhik ,"Table of Intergals, Series and Products." Academic Press. New York 1980.
\item    W. H. Huang, Quantization of Semiclassical Gravitation Theory in Anisotropic Spacetime, Phys. Letters. B 403, 8 (1997).

\end{enumerate}
\end{document}